# Composite fermion dynamics in half-filled Landau levels of graphene

ARKADIUSZ WÓJS[1,2], GUNNAR MÖLLER[1], NIGEL R. COOPER[1]

[1]TCM Group, Cavendish Laboratory, University of Cambridge, Cambridge, UK
[2]Institute of Physics, Wrocław University of Technology, Wrocław, Poland

We report on exact-diagonalization studies of correlated many-electron states in the half-filled Landau levels of graphene, including pseudospin (valley) degeneracy. We demonstrate that the polarized Fermi sea of non-interacting composite fermions remains stable against a pairing transition in the lowest two Landau levels. However, it undergoes spontaneous depolarization, which is unprotected owing to the lack of single-particle pseudospin splitting. These results suggest the absence of the Pfaffian phase in graphene.

## 1. Introduction

The continued interest in the physics of a half-filled Landau level (LL) is motivated by its mapping onto the system of composite fermions (CFs) [1] in zero effective magnetic field $B^*$ (the CFs are weakly interacting quasiparticles which experience a reduced effective field $B^*$, formed by the electrons through capturing vortices of the many-body wave function). At $B^*=0$, the non-interacting CFs form a Fermi sea. However, even weak attractive interactions among the CFs may lead to their pairing and to the formation of an incompressible liquid phase. The Moore-Read “Pfaffian” wave function [2] describing a paired CF liquid state supports quasiparticles with nonabelian braiding statistics [3], thus possibly opening the way to fault-protected topological quantum computation [4]. A wealth of numerical studies [5] have established the Pfaffian state as a promising candidate for the experimentally observed [6] fractional quantum Hall ground state in a half-filled second LL of conventional semiconductors, such as GaAs. However, the very small gap of this state in GaAs (hundreds of mK) [7] make its applications very challenging.

We investigate the possibility of this non-abelian phase appearing in half-filled LLs of graphene (characterized by a larger Coulomb energy scale than GaAs due to a lower dielectric constant and atomic thickness of a quasi-2D layer). Emergence of the Pfaffian phase requires a particular form of the interaction among the electrons inside a given LL – such as to support formation of CFs and to induce an interaction among the CFs themselves – such as to lead to their pairing at $B^*=0$. Hence, the question under consideration is that of CF dynamics in different LLs of graphene. The lack of immediate analogy with GaAs is a consequence of different single-particle orbitals defining all but the lowest LL, and of an additional “pseudospin” freedom associated with valley degeneracy in graphene. We find that, in this model, the CF-CF interaction in graphene is insufficient to induce pairing at $B^*=0$, and also that the polarized CF Fermi sea is unstable against depolarization of pseudospin.

## 2. Interaction pseudopotentials

Interaction within an isolated LL is defined by pseudopotentials $V_m$ (pair interaction energy as a function of relative angular momentum $m$). Except for the lowest LL, the pseudopotentials in graphene are different from those of GaAs due to the presence of a pair of atomic sublattices in its hexagonal crystalline structure. For the planar geometry, their analytic expressions have been derived earlier [8,9]. They were used in previous many-body calculations (e.g., [10]) carried out in a more convenient spherical geometry (with $N$ electrons confined to the surface of a sphere and exposed to a magnetic flux $2Q\ hc/e$), where the planar pseudopotentials for the $n$th LL were truncated at $m=2(Q+n)\equiv N_\phi$ (here, $N_\phi+1$ is the LL degeneracy; and $N_\phi$ stands for the magnetic flux of a corresponding system in the lowest LL).

The above approach disconnects $V_m$ from its Coulomb potential $V(r)\sim 1/r$ (especially at long range, comparable to the sphere radius, i.e., for $m$ approaching $Q$). Therefore, we instead have adopted a direct solution of the Dirac problem on a sphere [11]. In excited LLs of graphene, the single-particle states $\|n,m\rangle\rangle$ are spinors, whose two components represent standard LL wave functions, with equal angular momenta $m$ but different LL indices $n$ and $n$-1. It is essential that they must have equal $N_\phi$ (and thus different $2Q$). In this convention, the Coulomb matrix elements of graphene $\langle\langle\ldots\rangle\rangle$ are expressed as averages over

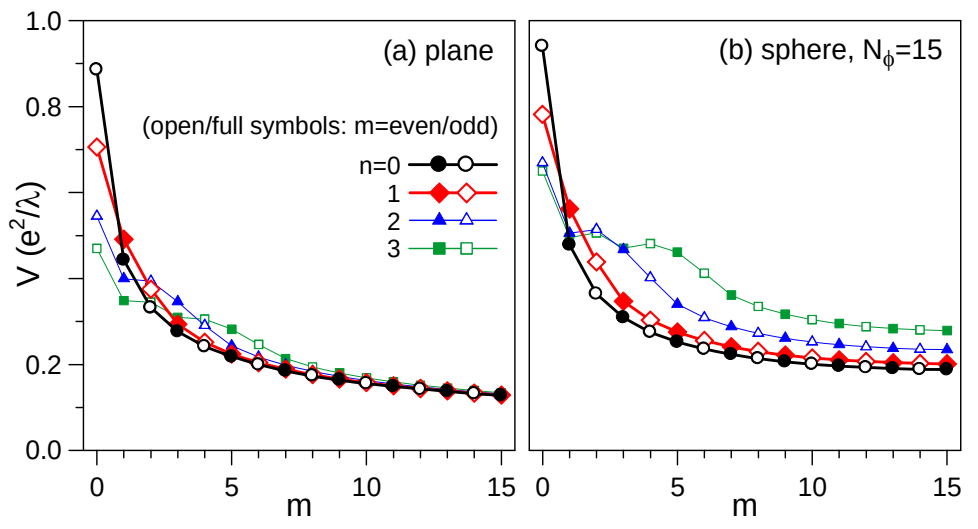

**Fig. 1** Coulomb pseudopotentials $V_m$ in the lowest LLs of graphene, calculated in the planar (a) and spherical (b) geometries; $\lambda=(hc/eB)^{1/2}$ is the magnetic length.

the corresponding two-body matrix elements of GaAs ⟨...⟩, all taken at the same $N_\phi$:

$$
\begin{aligned}
&4\langle\langle n,m_1;n,m_2|V|n,m_3;n,m_4\rangle\rangle \\
&\quad=\langle n,m_1;n,m_2|V|n,m_3;n,m_4\rangle \\
&\quad+\langle n-1,m_1;n,m_2|V|n-1,m_3;n,m_4\rangle \\
&\quad+\langle n,m_1;n-1,m_2|V|n,m_3;n-1,m_4\rangle \\
&\quad+\langle n-1,m_1;n-1,m_2|V|n-1,m_3;n-1,m_4\rangle.
\end{aligned}
\qquad (1)
$$

The pseudopotentials (describing graphene in the spherical geometry and correctly linked to the Coulomb potential $\sim 1/r$) shown in Fig. 1 are then obtained from the above matrix elements by diagonalization of the two-electron problem.

## 3. CF Fermi sea in polarized systems

Using matrix elements appropriate for the spherical geometry, we have first looked at the correlated $N$-electron states at $N_\phi=N/\nu-\sigma$ with the "filling factor" $\nu=1/2$ (corresponding to the half-filled LL) and various "shifts" $\sigma$. By analogy with a number of known cases, any extended incompressible liquid state is expected to be represented on a sphere by a series of uniform (i.e., having zero total angular momentum, $L$=0) finite-size states ($N$, $N_\phi$) with constant $\nu$ and $\sigma$ (the latter also deduced from the particular form of the extended many-body wave function, albeit in a less obvious way than $\nu$). Understandably, the correlation energy, pair correlation function, and all other features of the state should depend smoothly on $N$ along the series, and extrapolate to those describing an extended state on a plane.

We found that the nonabelian Pfaffian wave functions have only moderate overlaps with the corresponding ($\sigma$=3) exact $N$-electron Coulomb ground states in any LL of graphene. Moreover, their dependence on the system size $N$ reveals the emergence of a shell (CF-LL) structure of the (essentially) non-interacting CFs at $B^*$=0 rather than the formation of a uniform phase which could be adiabatically connected to the Pfaffian state, representing a paired CF liquid. This is illustrated in Fig. 2, showing also a comparison of the Coulomb energies with the Pfaffian states.

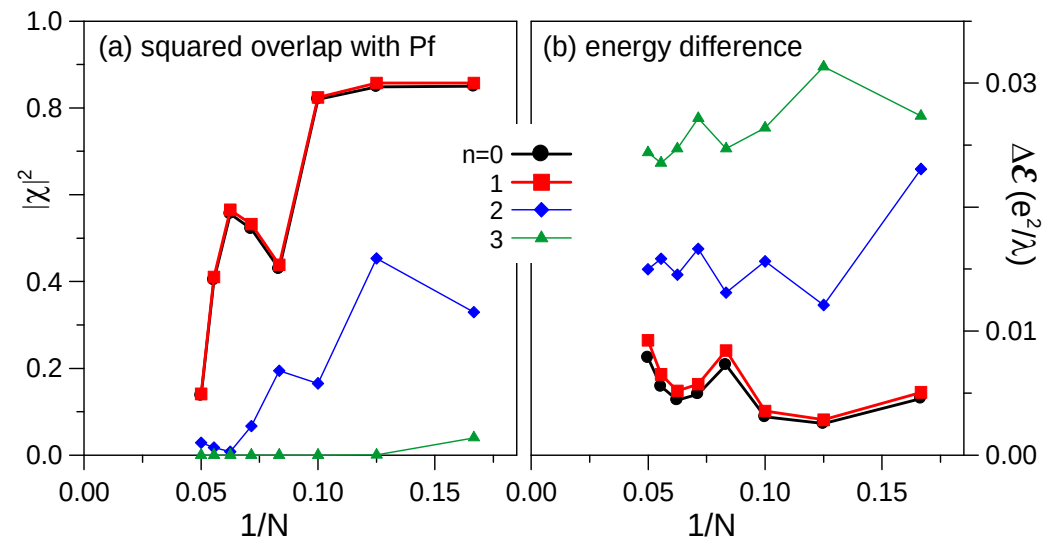

**Fig. 2** (a) Squared overlaps $|\chi|^2$ of the exact Pfaffian state with the lowest $L$=0 Coulomb eigenstate at the half-filling of different LLs in graphene (at the same shift $\sigma$=3), plotted as a function of an inverse electron number $N^{-1}$. (b) Difference $\Delta\mathcal{E}$ between the average Coulomb energy of the Pfaffian and the exact energy of the lowest $L$=0 Coulomb eigenstate, also at $\nu$=1/2 and $\sigma$=3, and for different LLs of graphene, versus $N^{-1}$.

The emergence of a shell structure is most evident in Fig. 3 showing the size dependence of the ground state correlation energy $E$ (counted per particle, including the charge compensating background, and found separately for each shift $\sigma$). The dominant tendency is the CF shell filling, with low values of $E$ coincident with exact filling of an indicated number of the CF-LLs.

Let us illustrate this tendency for a couple of examples. At $\nu$=1/2, the $(N,N_\phi)=(N,2N\text{-}\sigma)$ states of strongly interacting electrons map onto the non-interacting CF states at $N_\phi^*= N_\phi\text{-}2(N\text{-}1)=2\text{-}\sigma$ (the $N_\phi^*\sim N^0$ scaling being consistent with $B^*$=0). For $\sigma$=3, this gives $N_\phi^*$=-1. Negative sign means that the direction of an effective magnetic flux $2Q^*$, producing effective degeneracy $|N_\phi^*|+1=2$ of the lowest CF LL, is oriented opposite to the original magnetic flux $2Q$ [12]. The degeneracy of the $n^*$th CF-LL is $|N_\phi^*|+1+2n^*$. For $N$=20, four lowest CF-LLs are filled, and so this finite-size $\nu$=1/2 state is "aliased" with the incompressible $\nu^*$=-4 CF state, corresponding to the electron filling factor $\nu$=4/7 [12]. Hence, a relatively low correlation energy of this state is found in both frames of Fig. 3. Similarly, three lowest CF-LLs are full for $N$=12, this state being aliased with $\nu^*$=-3 (corresponding to $\nu$=3/5) and thus also having low energy in Fig. 3. For $\sigma$=4 (yielding $N_\phi^*$=-2) and $N$=16, there are three full CF-LLs and an additional single CF in the $n^*$=3 level with the single-particle angular momentum equal to $|N_\phi^*|/2+n^*$=4, yielding an $L$=4 state (aliased with the single CF-quasiparticle in an incompressible $\nu^*$=3 state) with a somewhat higher correlation energy than in the case of complete CF-LL filling. Finally, for $\sigma$=3 and $N$=16, there are four CFs in a half-filled $n^*$=3 CF-LL with angular momentum $|N_\phi^*|/2+n^*$=7/2. The four-CF spectrum depends on the form of residual CF-CF interaction; here, the lowest state has $L$=8 and a relatively high correlation energy in Fig. 3. These findings are consistent with the previous studies [8,10]. The second LL ($n$=1) behaves similarly to the lowest, also hosting a Fermi sea of nearly free CFs rather than an incompressible liquid of paired CFs.

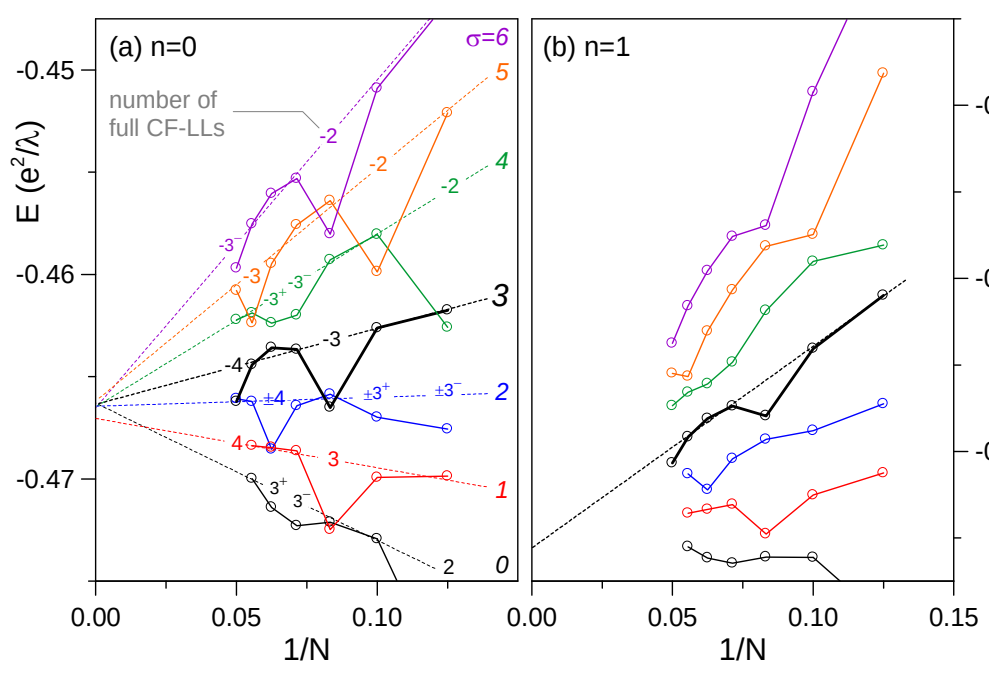

**Fig. 3** Correlation energies per particle in $N$-electron ground states corresponding to the half-filling of the lowest (a) and second (b) LL in graphene. Shift σ and the number of filled CF-LLs are explained in the text.

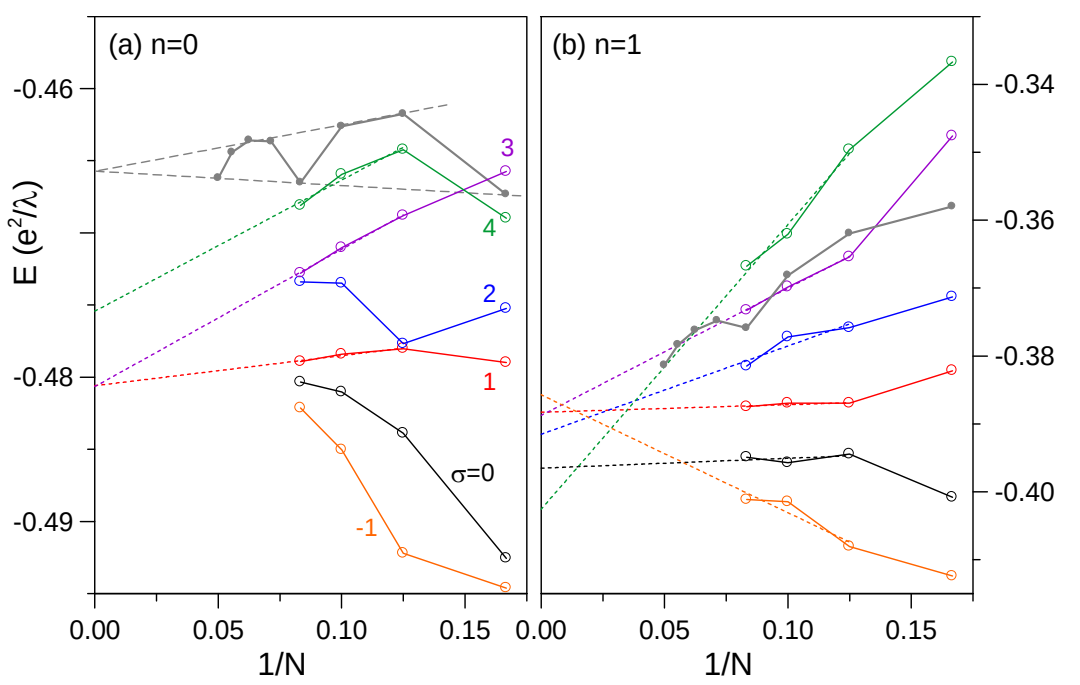

**Fig. 4** Similar to Fig. 2, but for the spin-unpolarized $N$-electron ground states at the half-filling in the lowest two LLs of graphene (for comparison, the polarized σ=3 series is also shown with grey symbols).

## 4. Pseudospin depolarization

Spin polarization at ν=1/2 (or in any other quantum Hall state in a high magnetic field) can in principle be enforced by a Zeeman splitting. However, in graphene one must also consider an additional "pseudospin" degree of freedom associated with the valley degeneracy. In fact, inclusion of the pseudospin in an analysis of the low-energy dynamics in graphene is of crucial importance as its depolarization is not protected by a single-particle splitting (in the absence of external potentials, or lattice distortions, that might break the valley degeneracy).

Hence, we have included the single-electron pseudospin ($p_z$=±1/2) in our diagonalization and classified the $N$-electron eigenstates by the total pseudospin $P$ and its projection $P_z$. Dependence of the correlation energy $E$ on size $N$, analogous to Fig. 3 but for the unpolarized states ($P$=0), is shown in Fig. 4. Especially for $n$=0 it is evident that a half-filled state undergoes a spontaneous pseudospin depolarization (e.g., the unpolarized series with σ=1, 3, and 4 clearly extrapolate to lower energies than any polarized series).

## 5. Conclusion

Using exact numerical diagonalization we have studied the correlated many-electron states in different half-filled LLs of graphene, including pseudospin (valley) degeneracy. We have found that even assuming the full spin and pseudospin polarization, the nonabelian Pfaffian state is not realized in graphene (at a half-filling of any LL). Instead, the essentially free CFs would form a Fermi sea in both lowest LLs if full polarization could be enforced (in higher LLs a striped order is most likely, but discussion of this issue has not been included). We also found that the half-filled ground states in both lowest two LLs undergo spontaneous depolarization of the pseudospin, which cannot be protected by a single-electron splitting (analogous to the Zeeman effect for spin). Together, our results point to the absence of a nonabelian Pfaffian phase in graphene.

We thank N. Regnault and M. O. Goerbig for helpful discussions. AW acknowledges support by the Marie Curie Grant PIEF-GA-2008-221701.

## References

[1] J. K. Jain, *Phys. Rev. Lett.* **63**, 199 (1989).
[2] G. Moore and N. Read, *Nucl. Phys. B* **360**, 362 (1991).
[3] C. Nayak and F. Wilczek, *Nucl. Phys. B* **479**, 529 (1996).
[4] C. Nayak et al., *Rev. Mod. Phys.* **80**, 1083 (2008).
[5] R. H. Morf, *Phys. Rev. Lett.* **80**, 1505 (1998); E. H. Rezayi and F. D. M. Haldane, *Phys. Rev. Lett.* **84**, 4685 (2000); G. Möller and S. H. Simon, *Phys. Rev. B* **77**, 075319 (2008); A. Wójs, *Phys. Rev. B* **78**, 041104(R) (2008); M. Storni, R. H. Morf, and S. Das Sarma, *Phys. Rev. Lett.* **104**, 076803 (2010); A. Wójs, G. Möller, S. H. Simon, and N. R. Cooper, *Phys. Rev. Lett.* **104**, 086801 (2010).
[6] R. L. Willett et al., *Phys. Rev. Lett.* **59**, 1776 (1987).
[7] C. R. Dean et al., *Phys. Rev. Lett.* **100**, 146803 (2008).
[8] M. O. Goerbig, R. Moessner, and B. Douçot, *Phys. Rev. B* **74**, 161407(R) (2006).
[9] K. Nomura and A. H. MacDonald, *Phys. Rev. Lett.* **96**, 256602 (2006).
[10] C. Tőke and J. K. Jain, *Phys. Rev. B* **76**, 081403(R) (2007).
[11] A. Jellal, *Nucl. Phys. B* **804**, 361 (2008).
[12] G. Möller and S. H. Simon, *Phys. Rev. B* **72**, 045344 (2005).